\documentclass[aps,prb,preprint,superscriptaddress,showpacs]{revtex4-1}

\usepackage[dvipdfm,colorlinks=true,linkcolor=blue,citecolor=blue,anchorcolor=blue,urlcolor=blue]{hyperref}
\usepackage{graphicx}

\begin{document}


\title{Enhanced weak-ferromagnetism and conductivity in hole-doped pyrochlore iridate Y$_2$Ir$_2$O$_7$}


\author{W. K. Zhu}
\author{M. Wang}
\author{B. Seradjeh}
\affiliation{Department of Physics, Indiana University, Bloomington, Indiana 47405}
\author{Fengyuan Yang}
\affiliation{Department of Physics, Ohio State University, Columbus, Ohio 43210}
\author{S. X. Zhang}
\email[]{sxzhang@indiana.edu}
\affiliation{Department of Physics, Indiana University, Bloomington, Indiana 47405}



\begin{abstract}
Pyrochlore iridates have recently attracted growing interest in condensed matter physics because of their potential for realizing new topological states. In order to achieve such quantum states, it is essential to understand the magnetic properties of these compounds, as their electronic structures are strongly coupled with their magnetic ground states. In this work, we report a systematic study of the magnetic properties of pyrochlore Y$_2$Ir$_2$O$_7$ and its hole-doped compounds by performing magnetic, electron spin resonance (ESR), electrical transport and x-ray photoelectron spectroscopy (XPS) measurements. We demonstrate the existence of weak ferromagnetism on top of a large antiferromagnetic background in the undoped compound. Hole-doping by calcium was found to enhance both the ferromagnetism and the electrical conductivity. The XPS characterization shows the coexistence of Ir$^{4+}$ and Ir$^{5+}$ in the undoped compound, and the amount of Ir$^{5+}$ increases with Ca-doping, which highlights the possible origins of the weak ferromagnetism associated with the formation of Ir$^{5+}$. We also observe a vertical shift in the $M$-$H$ curves after field cooling, which may arise from a strong coupling between the ferromagnetic phase and the antiferromagnetic background.
\end{abstract}


\maketitle

\section{INTRODUCTION}

5d transition metal oxides (TMOs) provide a fascinating system to study the interplay and competition between spin-orbit coupling (SOC) and electron correlation\cite{1,2,3,4,5,6} that have comparable energy scales.\cite{1,7} Among all the 5d TMOs, the pyrochlore iridates A$_2$Ir$_2$O$_7$ (A-227, A = Y, Bi, or lanthanide element) have attracted particular interests \cite{8,9,10,11,12,13,14,15,16,17,18,19,20,21,22,23,24,25,26,27,28,29} because of the recent theoretical predictions of novel topological phases.\cite{2,3,4} Indeed, a number of topologically non-trivial states, such as topological Mott insulators,\cite{2,3,30} Weyl semimetals,\cite{4,31} axion insulators \cite{4,32} and topological crystalline insulators,\cite{33} have been predicted to exist in certain correlation regimes that can potentially be accessed by tuning the A-site elements.\cite{6,34}

To realize novel topological states in pyrochlore iridates, it is essential to understand their magnetic properties as their electronic structures are strongly coupled with the magnetic ground states.\cite{4,14,31} In the pyrochlore lattice, the magnetic Ir$^{4+}$ ions form a corner-sharing tetrahedral network, and the competition between the three major magnetic interactions (i.e. the Heisenberg-type antiferromagnetic coupling, the Dzyaloshinskii-Moriya interaction, and the single-ion anisotropy)\cite{4,35,36} gives rise to a variety of magnetic configurations. When the A$^{3+}$ ion is magnetic, the possible \textit{f-d} exchange interaction\cite{14} between the Ir$^{4+}$ and A$^{3+}$ leads to even more complex magnetic structures.\cite{9,17,23,25,37,38} Therefore, to clarify the fundamental magnetic properties associated with the iridium tetrahedral network, it is desired to choose the compounds with non-magnetic A-site ions, e.g. Eu$^{3+}$,\cite{18,19} Bi$^{3+}$,\cite{26,27,28,29} and Y$^{3+}$.\cite{20,21,39,40} The Y$_2$Ir$_2$O$_7$ (i.e. Y-227) is of particular interest because it is predicted to be a Weyl semi-metal when its magnetic ground sate is the peculiar all-in/all-out antiferromagnetic (AFM) phase.\cite{4} Neutron scattering measurements do not show clear evidence of long-range magnetic order,\cite{20,21} which may be due to the low scattering intensity arising from the large neutron absorption cross-section of iridium. Muon spin rotation and relaxation experiments yield a well-defined muon spin precession frequency, indicating a long-range magnetic order at low temperatures.\cite{21} Further analysis of the spontaneous muon spin precession frequency in combination with the probabilistic and \textit{ab-initio} modeling techniques suggest that the all-in/all-out AFM is indeed the magnetic ground state.\cite{40} Despite these advances, the precise nature of the magnetism in this compound is far from being fully understood. For example, a small hysteresis loop was observed in the magnetization versus magnetic field ($M$-$H$) curve taken at low temperatures,\cite{20} which cannot be explained by the perfect all-in/all-out antiferromagnetic scenario.

In this work, we have performed systematic magnetization measurements, electron spin resonance (ESR) studies, transport characterizations and x-ray photoelectron spectroscopy measurements (XPS) to probe the magnetic properties of undoped and Ca-doped Y$_2$Ir$_2$O$_7$ compounds. We demonstrate the existence of weak ferromagnetism (FM) in the undoped sample, and the enhancement of both FM and electrical conductivity by Ca-doping. We also provide evidence of strong exchange coupling between the FM and the large AFM background. The XPS studies show mixed valence states of Ir, i.e. 0, 4+ and 5+, in both the undoped and doped samples.

\section{EXPERIMENTAL METHODS}

The (Y$_{1-x}$Ca$_x$)$_2$Ir$_2$O$_7$ samples were prepared by conventional solid state reaction. Mixtures of high purity Y$_2$O$_3$ (99.99\%), CaCO$_3$ (99.997\%) and IrO$_2$ (99.99\%) in stoichiometric ratios were heated in air at temperatures between 900 $^\circ$C and 1050 $^\circ$C for about 6 days with several intermediate grindings. The structure and phase purity were characterized by X-ray powder diffraction using a PANalytical EMPYREAN diffractometer (Cu K$\alpha$ radiation). The samples have a nearly pure phase of cubic \textit{Fd-3m} (227) except for some minor impurity phases of the source materials (see Fig. S1 in Supplemental Material),\cite{41} which is consistent with what has been reported by other groups.\cite{20,21} These impurity phases are either paramagnetic or diamagnetic, and hence have negligible contribution to the weak ferromagnetic properties that will be discussed below. The magnetization measurements were performed in a Quantum Design Magnetic Property Measurement System. The ESR spectra were taken in in a BRUKER EMX plus spectrometer with X band microwave (frequency $\nu$ = 9.40 GHz), following either zero-field cooling (ZFC) or field cooling (FC). The resistivity was measured using a Linear Research LR-700 AC Resistance Bridge. The XPS measurement was carried out in a PHI VersaProbe II Scanning XPS Microprobe and the fitting to the spectra was done using a standard software package CasaXPS provided by Casa Software Ltd.

\section{RESULTS AND DISCUSSION}

\begin{figure}[htbp]
\includegraphics[width=85mm]{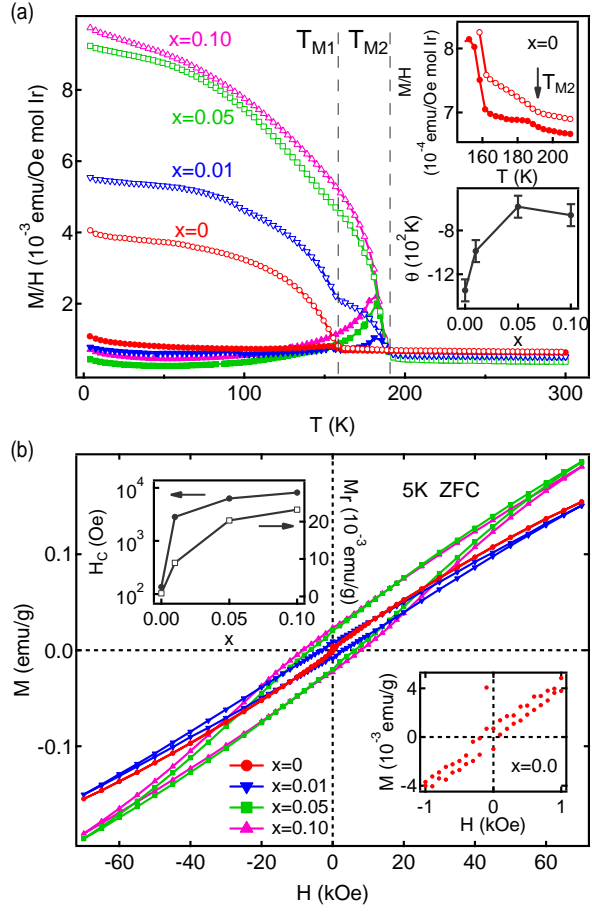}
\caption{(a) Temperature dependence of ZFC and FC susceptibilities of (Y$_{1-x}$Ca$_x$)$_2$Ir$_2$O$_7$ samples measured in a magnetic field $H$ = 1 kOe. The cooling field $H_{FC}$ = 1 kOe. Upper inset: magnetic susceptibilities of the undoped sample near 190 K; Lower inset: the Weiss constant as a function of doping concentration $x$.  (b) Magnetic hysteresis loops ($M$-$H$) taken at 5 K after ZFC. Upper inset: the coercivity $H_C$ and remanent $M_r$ versus doping concentration $x$; Lower inset: low field $M$-$H$ loop of undoped sample at 5 K.\label{f1}}
\end{figure}

The magnetic susceptibilities of (Y$_{1-x}$Ca$_x$)$_2$Ir$_2$O$_7$ ($x$=0, 0.01, 0.05 and 0.10) samples taken after ZFC and FC are plotted as a function of temperature in Fig. \ref{f1}(a). For the undoped sample ($x$=0), a clear magnetic transition is observed at $T_{M1}$ $\sim$ 158 K, below which there is a large hysteresis difference between the ZFC and FC magnetic susceptibilities. This result is consistent with earlier studies,\cite{20,21,37,39,42} confirming our sample quality. We also noticed that there is a very small but visible kink around 190 K (upper inset of Fig. \ref{f1}(a)), indicating a possible transition. A more pronounced transition was observed by Disseler et al.\cite{21} With Ca-doping, the FC magnetization is enhanced and the transition at $T_{M2}$ $\sim$ 190 K becomes more prominent. For higher doping level of $x$=0.05 and 0.10, the transition at $T_{M2}$ dominates over the one at $T_{M1}$. By fitting the magnetic susceptibility data above $T_{M2}$, we obtain the Weiss constant $\theta$ as shown in the lower inset of Fig. \ref{f1}(a). All the Weiss constants are negative with unusually large absolute values, suggesting a strong AFM interaction in all the samples. With Ca-doping, the constant increases, which indicates the weakening of AFM interaction.

While the AFM interaction and the small magnitude of magnetization ($\sim$ 10$^{-4}$ $\mu_B$/Ir) are consistent with the all-in/all-out configuration, the $M$-$H$ data taken at 5 K show hysteresis loops (Fig. \ref{f1}(b)), which suggests  the existence of ferromagnetic component on top of the AFM background. We plot the magnetic coercivity $H_C$ and the remnant magnetization $M_r$ as a function of doping concentration in the upper inset of Fig. \ref{f1}(b). As the magnetic hysteresis loops of the $x$=0.05 and 0.10 samples are not saturated and there is a large AFM background for all the samples, their $H_C$ and $M_r$ values may be underestimated. Nevertheless, these two parameters show clear increase with increasing the doping concentration, suggesting the enhancement of ferromagnetism by doping.

\begin{figure}[htbp]
\includegraphics[width=85mm]{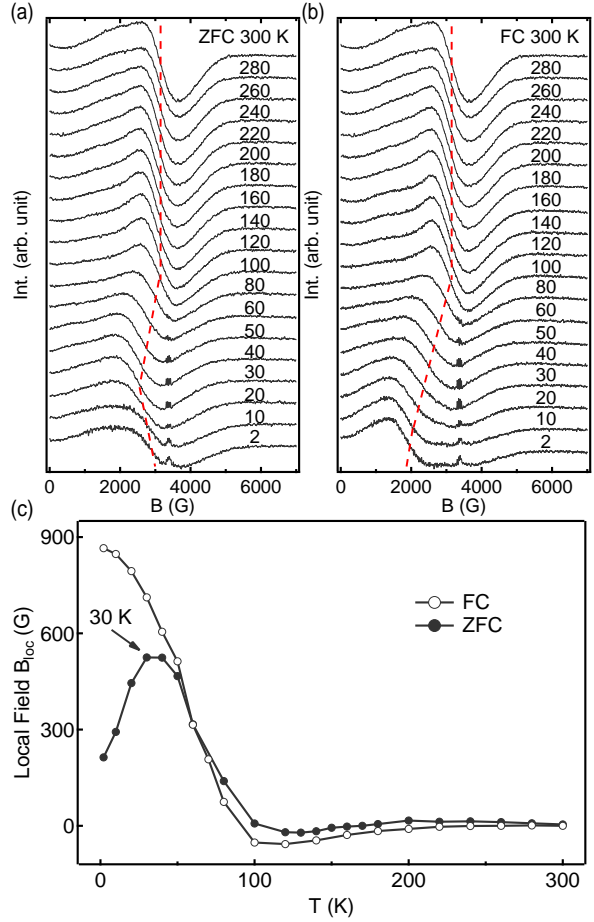}
\caption{ESR spectra of undoped Y$_2$Ir$_2$O$_7$ at different temperatures after (a) ZFC and (b) FC ($H_{FC}$ = 1 kOe). The spectra are shifted vertically for clarity. The red dashed lines indicate the location of resonant fields. (c) Temperature dependence of effective local field $B_{loc}$.\label{f2}}
\end{figure}

Since the magnetic hysteresis loop of the undoped sample is small (lower inset of Fig. \ref{f1}(b)), we further performed ESR measurements to confirm the weak-ferromagnetism. Fig. \ref{f2}(a) and (b) show the ESR spectra (i.e. first derivative of absorption with respect to field) of the undoped sample taken at variable temperatures after the sample was cooled down in zero field and in a field of 1 kOe, respectively. To obtain the resonant field $B_r$, we fit the absorption spectra using the Gaussian function plus a polynomial function which represents the background (see Fig. S2 in Supplemental Material).\cite{41} The resonant field was found to be $\sim$ 3050 G at 300 K where the sample is in a paramagnetic state, and it almost remains the same above 100 K (Fig. \ref{f2}(a) and (b)). When the temperature is below 100 K, it shifts towards lower values, which suggests the appearance of an effective local magnetic field that is created by the ferromagnetically ordered moments. The resonant field decreases until 30 K, after which the ZFC and FC curves behave differently. The ZFC resonant field has a minimum value at 30 K, and then increases with the decrease of temperature. In contrast, the FC resonant field decreases monotonically with decreasing temperature towards 2 K. We note that recently Liu et al. also reported a small shift of resonant field in their ESR measurements,\cite{43} but their data were taken under ZFC with less clear temperature dependence.

We calculated the effective local field $B_{loc}$($T$) = $B_r$(PM) - $B_r$($T$), where $B_r$(PM) = 3050 G is the resonant field at 300 K (i.e. in the paramagnetic regime) and $B_r$($T$) represents the resonant field at a temperature $T$. Fig. \ref{f2}(c) shows the temperature dependence of local field. Upon warming, the ZFC local field increases from $\sim$ 210 G at 2 K to $\sim$ 520 G at 30 K. It then decreases with further increase of temperature and reaches zero above 100 K. The FC local field is $\sim$ 870 G at 2 K, and it decreases monotonically with increasing temperature. The different temperature dependence between ZFC and FC should be attributed to the freezing of ferromagnetic domains at low temperatures. In brief, the FM domains are frozen in nearly random directions after the sample was cooled down in zero field, while they are aligned along the field direction after FC. As a result, the effective local field is lower in the ZFC measurement than in the FC. We also note that although the non-zero local field is an indication of ferromagnetism, weak or short-range ordered FM may not produce a large enough local field that can be detected by ESR,\cite{44,45} Therefore, we cannot draw the conclusion that the ferromagnetic phase only exists below 100 K. In fact, as we will discuss later the ferromagnetic phase persists up to $T_{M2}$ $\sim$ 190 K. In addition to the FM component, we also observed a hyperfine resonance at a higher field ($\sim$ 3350 G) (see Fig. S3 in Supplemental Material),\cite{41} which suggests the existence of diluted and isolated paramagnetic moments besides the strong AFM and weak FM phases. This paramagnetic phase is consistent with the sharp increase of magnetic susceptibility upon cooling at low temperatures (Fig. \ref{f1}(a)).

\begin{figure}[htbp]
\includegraphics[width=85mm]{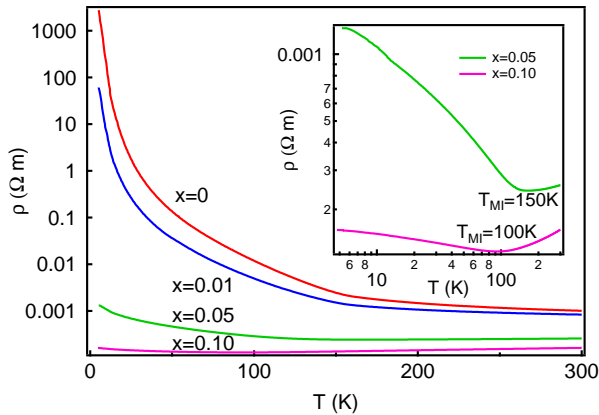}
\caption{Temperature dependence of electrical resistivity for all (Y$_{1-x}$Ca$_x$)$_2$Ir$_2$O$_7$ samples. Inset: magnification of the data for the $x$=0.05 and 0.10 samples.\label{f5}}
\end{figure}

The ESR spectra taken on the Ca-doped samples do not show a visible resonance response near 3050 G. We note that ESR mainly detects the local environment of the localized unpaired electrons. The absence of resonance could be due to the decrease and/or the delocalization of unpaired electrons arising from Ca-doping: first, the substitution of Y$^{3+}$ by Ca$^{2+}$ is expected to change Ir$^{4+}$ to Ir$^{5+}$ that does not have the unpaired electron; second, this doping could lead to the delocalization of the unpaired electron of Ir$^{4+}$, as evidenced by the significant decrease of electrical resistivity (Fig. \ref{f5}).

\begin{figure}[htbp]
\includegraphics[width=85mm]{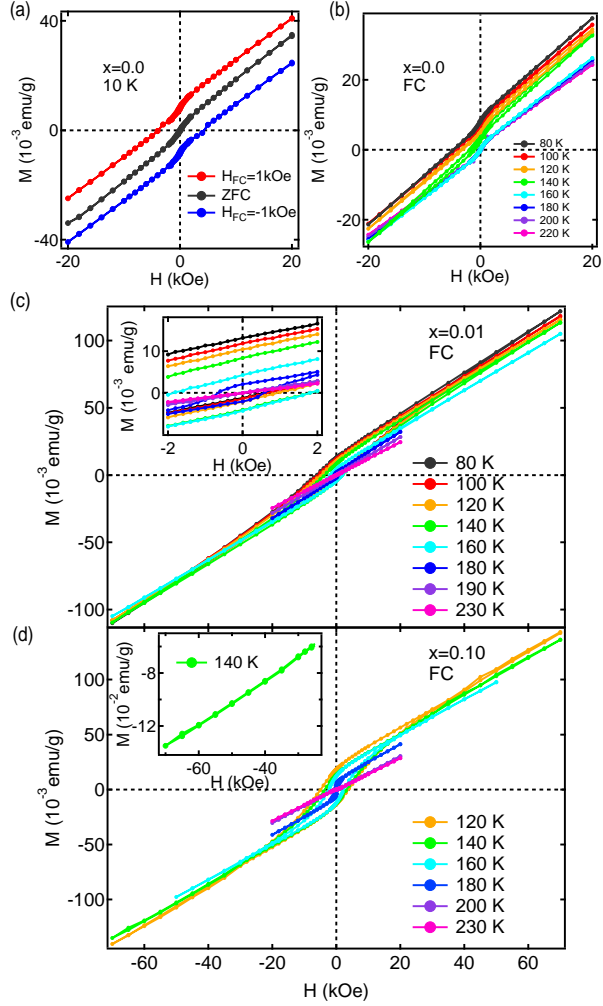}
\caption{(a) $M$-$H$ curves of undoped Y$_2$Ir$_2$O$_7$ taken at 10 K after ZFC and FC ($H_{FC}$ = 1 kOe, -1 kOe). (b) $M$-$H$ curves of undoped sample taken at variable temperatures near $T_{M1}$ and $T_{M2}$. $M$-$H$ curves of (c) $x$=0.01 and (d) $x$=0.10 samples taken after FC ($H_{FC}$ = 1 kOe). The inset of (c) shows the $M$-$H$ curves at low field; the inset of (d) shows a typical $M$-$H$ curve at high field.\label{f3}}
\end{figure}

We further performed $M$-$H$ measurements after the samples were cooled down in a magnetic field to study the possible exchange coupling between the ferromagnetism and the antiferromagnetic background. Fig. \ref{f3}(a) shows three $M$-$H$ curves that were taken after the undoped sample was cooled down to 10 K in 1 kOe, 0 and -1 kOe, respectively. The magnetic field was swept between -2 T and 2 T, which is high enough to saturate the hysteresis loop. While the ZFC $M$-$H$ curve passes through the origin, the two FC curves show clear shift along the vertical axis. In particular, a positive cooling field leads to a positive shift while a negative cooling field gives rise to a negative shift. We noticed that a similar vertical shift was observed in the Sm$_{2}$Ir$_{2}$O$_{7}$ compound,\cite{37} but its underlying mechanism was unclear. Based on the above analysis regarding the coexistence of FM and AFM phases, we propose that the strong exchange coupling at the interface between these two separated phases is a possible origin of the shift. As demonstrated in some FM-AFM coupled systems \cite{46,47} including the phase-separated manganites,\cite{48} the magnetic moments in the shell of the FM phase (i.e. at the interface between FM and AFM domains) are strongly pinned by the exchange coupling from the AFM phase. The magnetic moments in the FM domains are aligned with the cooling field, which gives rise to a net moment when the magnetic field is removed.  When the magnetic field is swept from 2 T to -2 T, the negative field is able to align the moment in the bulk FM, but not strong enough to rotate the moments in the shell of the FM that is pinned by the AFM phase. These uncompensated moments hence give rise to a vertical shift of the hysteresis loop. We note that in conventional exchange bias systems\cite{47} the magnetic hysteresis loop is shifted horizontally. The vertical shift observed here (similar to the observation in manganites\cite{48}) suggests that the FM and AFM phases are strongly coupled, resulting in a strong pinning of interfacial moments which cannot be rotated/aligned by the sweeping field.

\begin{figure}[htbp]
\includegraphics[width=85mm]{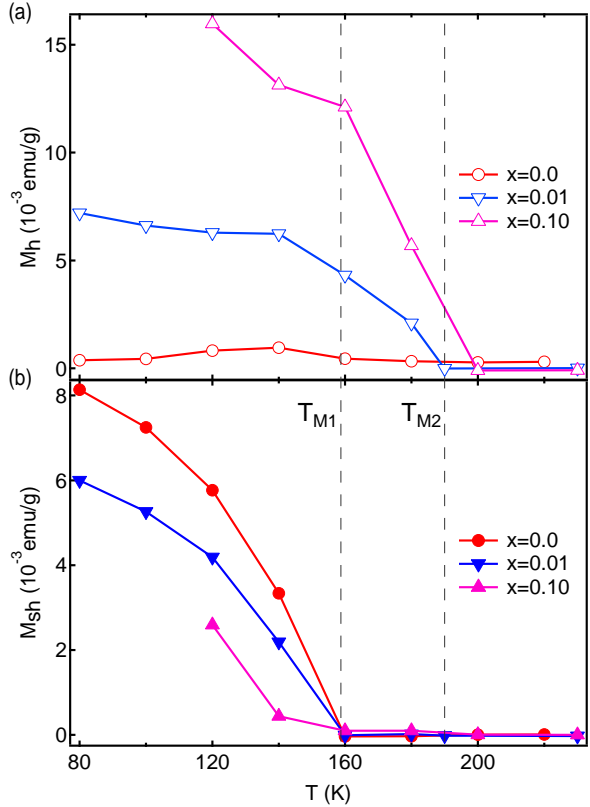}
\caption{(a) The hysteresis $M_h$ and (b) the shifted magnetization $M_{sh}$ as a function of temperature for the $x$=0.0, 0.01 and 0.10 samples.\label{f4}}
\end{figure}

We carried out more FC $M$-$H$ measurements at various temperatures near $T_{M1}$ $\sim$ 158 K and $T_{M2}$ $\sim$ 190 K for the $x$=0, 0.01 and 0.10 samples. At low temperatures, the magnetic field is swept between -7 T and 7 T for the doped samples to ensure that their magnetic hysteresis loops are saturated (inset of Fig. \ref{f3}(d)). As shown in Fig. \ref{f3}(b), (c) and (d), all the hysteresis loops show vertical shift below $T_{M1}$. We plot the temperature dependence of the hysteresis $M_h$ and shift $M_{sh}$ in Fig. \ref{f4}(a) and (b), respectively. Here the hysteresis $M_h$ is defined as one half of the difference between the two magnetization values at zero field, and the shift $M_{sh}$ is defined as the average of these two values. It is clear that the $M_{sh}$ reaches zero above $T_{M1}$ $\sim$ 158 K while the $M_h$ persists up to about $T_{M2}$ $\sim$ 190 K. These results suggest that the weak ferromagnetism exists up to $T_{M2}$, while the coupling between FM and AFM occurs below $T_{M1}$. This is consistent with the recent muon spin resonance measurement which suggests that the all-in/all-out AFM ordering is formed below $\sim$ 150 K in the undoped Y-227.\cite{21} At each temperature, $M_h$ increases with doping concentration, consistent with the enhancement of ferromagnetism. On the other hand, the $M_{sh}$ which is related to the pinned magnetic moments decreases with doping. This suggests that the interfacial coupling between the FM and AFM becomes weaker. We note that recent $\mu$SR measurement on the undoped sample does not seem to show signature of FM phase,\cite{21} which could be due to the small faction of FM component. Indeed a rough estimate based on the magnetic hysteresis loop suggests that the volume fraction of the FM phase is less than 0.1\% for the undoped sample, which may not give rise to a detectable signal in $\mu$SR.

We further performed electrical transport studies to gain insight into the nature of the observed weak-ferromagnetism. Fig. \ref{f5} shows the temperature dependence of electrical resistivity for all four samples. The undoped sample shows an insulating behavior and the $\rho$-$T$ curve has a kink at the magnetic transition $T_{M1}$ $\sim$ 158 K, consistent with the previous reports.\cite{43} The Ca-doping significantly enhances the electrical conductivity (i.e. reduces the resistivity). In particular, the $x$=0 and 0.01 samples are insulators, while the $x$=0.05 and 0.10 samples show a metal-insulator (MI) transition at about 150 K and 100 K, respectively.

The doping of Ca$^{2+}$ has two effects. First, it increases the A-site ionic radius, which could enhance the conductivity by reducing the trigonal compression on the IrO$_6$ octahedra.\cite{6,49} Indeed, earlier studies have shown that, upon the increase of the A-site radius, the transport properties of A-227 change successively from insulating behavior,\cite{21,34,39,50,51} to MI transition,\cite{12,13,23,34,51} and finally to metallic behavior.\cite{26,29,34,50,51} Ca$^{2+}$ has an ionic radius of 1.12 {\AA}\cite{52} which is slightly larger than that of Y$^{3+}$ (1.019 {\AA}).\cite{52} With a doping of $x$=0.05, the average A-site radius is increased to 1.024 {\AA}, which is slightly smaller than that of Gd$^{3+}$ (1.053 {\AA}).\cite{52} We note that Gd-227 is still insulating over the entire temperature range (5 - 300 K),\cite{34,50} so the high temperature metallic behavior of our Ca-doped samples is unlikely to be due to the increase of A-site radius. The second effect of Ca-doping is the inducing of charge carriers. As discussed earlier, the substitution of Y$^{3+}$ by Ca$^{2+}$ increases the valence state of Ir from 4+ to 5+.  In the stoichiometric A-227 compounds, the Ir$^{4+}$ has an unpaired $J_{eff}$=1/2 electron that is localized due to the electron-electron interaction.\cite{1,2,4} In the doped compounds, the Ir$^{5+}$ has an empty $J_{eff}$=1/2 level which allows the hopping of the $J_{eff}$=1/2 electron from the nearby Ir$^{4+}$, leading to the delocalization of electrons and enhancement of electrical conductivity.

To characterize the valence state of Ir, we performed XPS measurements on the $x$=0 and 0.10 samples. A set of Ir 4$f$ spectra are shown in Fig. \ref{f6} and another set of spectra taken at a different depth are shown in Fig. S4 in Supplemental Material to demonstrate the consistency of spectra through depth. The apparent feature of multi-peaks indicates the coexistence of different valence states of iridium for both samples. We fitted the spectra using a standard software package called CasaXPS. It is worth noting that the screening effect observed in the metallic IrO$_2$\cite{55} is negligible in our samples due to their much higher resistivities,\cite{55,56} -therefore all the fitting components are considered to be symmetric with a Voigt shape. As shown in Fig. \ref{f6}, the experimental data can be fitted well using three sets of iridium components. Taking the 4$f_{7/2}$ spectra as an example, the component with the lowest binding energy of 60.9 eV (blue) is determined to be Ir,\cite{53} consistent with the existence of Ir impurity in the XRD pattern (see Fig. S1 in Supplemental Material).\cite{41} The intermediate component (62.4 eV, red) is Ir$^{4+}$\cite{54} which is the nominal valence state of Ir in the undoped sample. The component with the highest binding energy of 65 eV (green) should be attributed to a valence state that is higher than 4+, i.e. 5+ or 6+.  Since this high valence component is found to increase by $\sim$ 8\% with $\sim$ 10\% doping of Ca$^{2+}$ , it is likely to be Ir$^{5+}$ instead of Ir$^{6+}$. Moreover, as discussed above, the coexistence of Ir$^{5+}$ and Ir$^{4+}$ can give rise to delocalization of $J_{eff}$=1/2 electron and enhancement of electrical conductivity, while the same effect is not applicable for a mixed valence states of 6+ and 4+. The existence of Ir$^{5+}$ in the undoped sample should be attributed to non-stoichiometry, e.g. deficiency of metal elements (Y or Ir) or excess oxygen.

\begin{figure}[htbp]
\includegraphics[width=85mm]{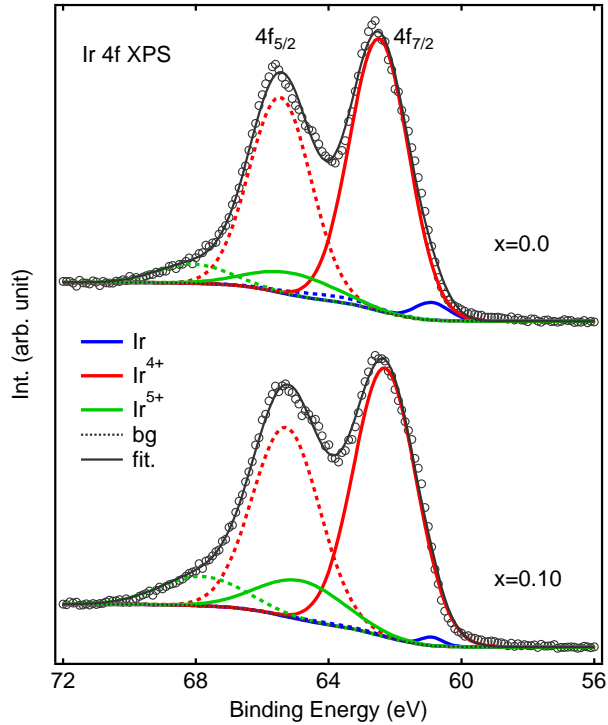}
\caption{Typical Ir 4$f$ XPS spectra of Y$_2$Ir$_2$O$_7$ and (Y$_{0.9}$Ca$_{0.1}$)$_2$Ir$_2$O$_7$. The black solid curve is the fitted envelope using three individual components, i.e. Ir, Ir$^{4+}$ and Ir$^{5+}$ which are plotted in blue, red and green colors, respectively. The black dashed curve denotes the background.\label{f6}}
\end{figure}

Now we briefly discuss the possible origin of the observed ferromagnetism based on the known valence state of Ir. In the pyrochlore iridates, there are three major magnetic interactions, namely the Heisenberg AFM interaction, single-ion anisotropy and the D-M interaction.\cite{4,35,36} Theoretical calculations predicted that the interplay and competition among these three interactions stabilize the all-in/all-out antiferromagnetic phase.\cite{4} We note that these studies were based on the assumption that all the Ir are in the 4+ valence state. However, according to our XPS measurements, there are some Ir$^{5+}$ in both the undoped and doped samples. In contrast to Ir$^{4+}$ which has a magnetic moment of 1/3 $\mu_B$,\cite{20} the Ir$^{5+}$ has an empty $J_{eff}$=1/2 level along with fully occupied $J_{eff}$=3/2 states, and hence has no net moment. As a result, the replacement of Ir$^{4+}$ by Ir$^{5+}$ could modify the competition of those three magnetic interactions between the Ir$^{4+}$ ions in the vicinity of Ir$^{5+}$ and may favor a ferromagnetic state. Furthermore, the mixed-valence states of Ir may lead to a double-exchange interaction: the O2$p$ electron hops to the empty $J_{eff}$=1/2 orbital of Ir$^{5+}$, and then the $J_{eff}$=1/2 electron on the nearby Ir$^{4+}$ hops to the O2$p$ orbital, giving rise to electrical conductivity as discussed earlier. This double-exchange interaction between Ir$^{5+}$ and Ir$^{4+}$ through the oxygen 2$p$ orbital is similar to the Mn$^{3+}$-O-Mn$^{4+}$ interaction in the well-known manganites,\cite{57} and may give rise to ferromagnetism. We note that future theoretical calculations considering Ir$^{5+}$ will help understand the precise nature of the ferromagnetism in these compounds.

\section{CONCLUSIONS}

In this work, we performed systematic studies of the magnetic properties of a prototypical pyrochlore iridate Y$_2$Ir$_2$O$_7$ and its hole-doped compounds. We have demonstrated the existence of weak ferromagnetism in the undoped compound through a combination of magnetic characterizations and electron spin resonance studies. Ca-doping leads to the enhancement of ferromagnetism and improvement of electrical conductivity. We have also observed a vertical shift in the $M$-$H$ curves, which suggests a strong coupling between the ferromagnetic phase and the large antiferromagnetic background. The XPS characterization shows the existence of Ir$^{4+}$ and Ir$^{5+}$ in both the doped and undoped samples, and the amount of Ir$^{5+}$ increases with Ca-doping, which highlights the possible origins of the weak ferromagnetism associated with the formation of Ir$^{5+}$.

\begin{acknowledgments}
We thank Professors Herb Fertig, Gerardo Ortiz, Kai Sun and Luis Brey for helpful discussions, and Dr. D. Williams, W. Tong, L. Pi, and L. Ling for experimental assistance. S.X.Z. and B. S. would like to thank Indiana University (IU) College of Arts and Sciences for start-up support. F. Y. Y. acknowledges support from the Center for Emergent Materials at the Ohio State University, a NSF Materials Research Science and Engineering Center (DMR-1420451). We acknowledge the use of SQUID (partially) and ESR facilities in the High Magnetic Field Laboratory, Chinese Academy of Sciences at Hefei, and the use of x-ray diffraction and XPS instruments at the IU Molecular Structure Center.
\end{acknowledgments}


\end{document}